\documentclass[twocolumn]{jpsj2}

\def\runtitle{Effective Heisenberg-Model Description to the 
Coupled Spin-Pseudospin ...}
\def\runauthor{Tohru {\sc Nakaegawa} and Yukinori {\sc Ohta}}

\title{Effective Heisenberg-Model Description of the Coupled 
Spin-Pseudospin Model\\ for Quarter-Filled Ladders}

\author{Tohru {\sc Nakaegawa}$^{1}$ and Yukinori {\sc Ohta}$^{1,2}$
\thanks{E-mail: ohta@science.s.chiba-u.ac.jp}}

\inst{$^1$Graduate School of Science and Technology, Chiba 
University, Chiba 263-8522\\
$^2$Department of Physics, Chiba University, Chiba 263-8522}

\recdate{April 5, 2002}

\abst{Quantum Monte Carlo method is used to study the coupled 
spin-pseudospin Hamiltonian in one-dimension (1D) that models 
the charge-ordering instability of the anisotropic Hubbard 
ladder at quarter filling.  We calculate the temperature 
dependence of the uniform spin susceptibility and the 
spin and charge excitation spectra of the system to show 
that there is a parameter and temperature region where the 
spin degrees of freedom behave like a 1D antiferromagnetic 
Heisenberg model.  Anomalous spin dynamics in the 
disorder phase of a typical charge-ordered material 
$\alpha'$-NaV$_2$O$_5$ is thereby considered.}

\kword{charge ordering, charge fluctuation, Hubbard ladder, 
quarter filling, pseudospin}

\begin{document}
\sloppy
\maketitle

\section{Introduction}

Charge-ordering (CO) instability has recently been one 
of the major topics in the field of strongly correlated 
electron systems.  Here, elucidation of the observed 
anomalous behaviors of electrons associated with the CO 
phase transition has been the central issue.  This includes 
questions on the charge dynamics above the transition 
temperature $T_{\rm CO}$ as well as on the CO spatial 
patterns realized below $T_{\rm CO}$.  
A well-known example is the vanadate bronze 
$\alpha'$-NaV$_2$O$_5$ where the system may be modeled 
as a lattice of coupled ladders (or a trellis lattice) 
at quarter filling.\cite{smolinski,seo,nishimoto1,
mostovoy1,thalmeier}  
Strong intersite Coulomb interaction between electrons 
is believed to be the origin of the CO 
instability.\cite{seo,nishimoto1}  In this material, 
the CO with a zigzag ordering pattern is observed below 
$T_{\rm CO}=34$ K,\cite{isobe,ohama1,sawa,johnston,ohama2}  
and associated with this, a number of anomalous behaviors, 
which can be related to the slow dynamics of charge 
carriers (or charge fluctuation), have been observed 
above $T_{\rm CO}$.\cite{ravy,nakao,damascelli,presura,
ohama2,nishimoto2,nishimoto3,mostovoy2}  
Anomalous response of the spin degrees of freedom 
has also been noticed.\cite{hemberger,johnston,ohama3}
It seems therefore quite natural to wonder how in such 
systems the spin degrees of freedom behave near the CO 
phase transition when they are on the slowly fluctuating 
charge carriers.  
In this paper, we thus consider the issue: what are 
the consequences of charge fluctuation at $T>T_{\rm CO}$ 
to the spin degrees of freedom?  

One of the simplest models that allow for such situation 
is the anisotropic Hubbard ladders at quarter filling 
with the strong intersite Coulomb repulsion.   
We here use an effective Hamiltonian written in terms of 
the spin and pseudospin (representing charge degrees of 
freedom) operators.\cite{mostovoy1,cuoco,sa,mostovoy2}  
This Hamiltonian is derived from the Hubbard ladder model 
by the perturbation theory\cite{mostovoy1,cuoco,sa} where 
the hopping parameter between the rungs of the ladder is 
assumed to be small compared with the onsite and intersite 
Coulomb repulsions as well as the hopping parameter in the 
rung (i.e., the {\em anisotropic} ladder).\cite{nishimoto1}  
Although the CO is not realized in this model (since it 
is the 1D quantum-spin model), we can simulate anomalous 
behaviors of the spin degrees of freedom under the strong 
charge fluctuation.  
We will apply the quantum Monte Carlo (QMC) method to 
this model to calculate the temperature dependence of the 
uniform spin susceptibility and the spin and charge excitation 
spectra, thereby clarifying consequences of the interplay 
between its spin and charge degrees of freedom.  

We note that, in this coupled spin-pseudospin model, 
the spin exchange interaction is necessarily associated with 
the charge excitation; i.e., the spin excitations cannot 
occur without making the exchange of the pseudospins.  
We will then show that nevertheless there is a parameter 
and temperature region where the spin degrees of freedom 
behave like a 1D antiferromagnetic Heisenberg model; 
i.e., the spin degrees of freedom are `separated' from the 
charge degrees of freedom in this region.  
We will moreover show that the spin system behaves in different 
manner depending on whether the temperature $T$ is below or 
above a crossover temperature $T^*$ which is related to the 
pseudospin excitations; at $T\lesssim T^*$, it behaves like 
a 1D antiferromagnetic Heisenberg model with a $T$-independent 
effective exchange coupling constant $J_{\rm eff}$ with large 
renormalization, whereas at $T\gtrsim T^*$, $J_{\rm eff}$ 
decreases rapidly with increasing $T$, where the effective 
Heisenberg-model description ceases to be valid.  

This paper is organized as follows.  In \S2, we define 
the coupled spin-pseudospin model that describes the spin 
and charge degrees of freedom of the anisotropic Hubbard 
ladder at quarter filling.  Some details of the method 
of calculation are also given.  In \S3, we present the 
results of calculation which include the staggered 
susceptibility for pseudospins, the spin and pseudospin 
excitation spectra, and the temperature dependence of 
the uniform spin susceptibility.  Discussion on the 
experimental relevance to $\alpha'$-NaV$_2$O$_5$ and 
summary of the paper will be given in \S4.  

\section{Model and Method}

Our effective spin-pseudospin Hamiltonian may be written 
as a sum
\begin{equation}
{\cal H}={\cal H}_0+{\cal H}_{\rm ST}
\end{equation}
of the quantum Ising Hamiltonian for pseudospins 
\begin{equation}
{\cal H}_0=J_1\big(-\frac{g}{2}\sum_iT_i^x
+\sum_iT_i^zT_{i+1}^z\big)
\end{equation}
and the spin-pseudospin coupling term
\begin{equation}
{\cal H}_{\rm ST}=J_2\sum_i
\big({\bf S}_i\cdot{\bf S}_{i+1}-\frac{1}{4}\big)
\big(T_i^+T_{i+1}^-+{\rm H.c.}\big).  
\end{equation}
The standard notation is used here.  
${\bf S}_i$ and ${\bf T}_i$ are, respectively, the spin 
and pseudospin operators of spin-1/2 at site $i$, where 
$T_i^z=-1/2$ ($+1/2)$ means the electron is on the left 
(right) site on the rung of the ladder.  
$J_1$ is the energy scale of the pseudospin system and 
$J_2$ is the coupling strength between the spin and 
pseudospin systems.  

From the second-order perturbation 
theory,\cite{mostovoy1,cuoco,sa} we have the relations 
$J_1=2V_\parallel$ and $J_2=4t_\parallel^2/V_\perp$, 
where $t_\parallel$ and $V_\parallel$ ($t_\perp$ and 
$V_\perp$) are the nearest-neighbor hopping parameter 
and Coulomb repulsion of the leg (rung) of the ladder, 
respectively.  We should then have $J_1>J_2$, which 
we assume throughout the present work.  
We also assume the onsite Coulomb repulsion to be 
$U\rightarrow\infty$.  
Relative strength of the transverse field to the 
pseudospins is measured by 
$g=4t_\perp/J_1=2t_\perp/V_\parallel$.  
Note that $g$ in the quantum Ising model represents 
the relative strength of the fluctuation of a charge 
in the rung: if we assume one electron in a rung, we 
have the prefactor $gJ_1/2$ in the first term of eq.~(2), 
which is the difference between the energies of the 
bonding and antibonding levels of the rung, $2t_\perp$. 
Thus, if $g$ (or $t_\perp$) is large the electron is 
stable in the bonding level of the rung, but if $g$ 
(or $t_\perp$) is small the effect of $V_\parallel$ 
easily leads the system to CO.  

We use the conventional world-line QMC method for the 
analysis of the model.  We use a 32-site cluster (where 
a site contains a spin and a pseudospin) with 
periodic boundary condition; the cluster-size dependence 
of the calculated results are examined by using clusters 
up to 96 sites but we find no significant size 
dependence in the results.  
Because the model does not conserve the total pseudospin, 
we have examined a number of ways of the spin flips 
and confirmed that available analytical results are 
reproduced correctly.\cite{nakaegawa}
The maximum-entropy method is used to calculate 
the dynamical quantities like the spin and pseudospin 
excitation spectra.  

\section{Calculated Results}

\subsection{Staggered susceptibility for pseudospins}

\begin{figure}[h]
\vspace{5pt}
\begin{center}
\includegraphics[width=6.5cm,clip]{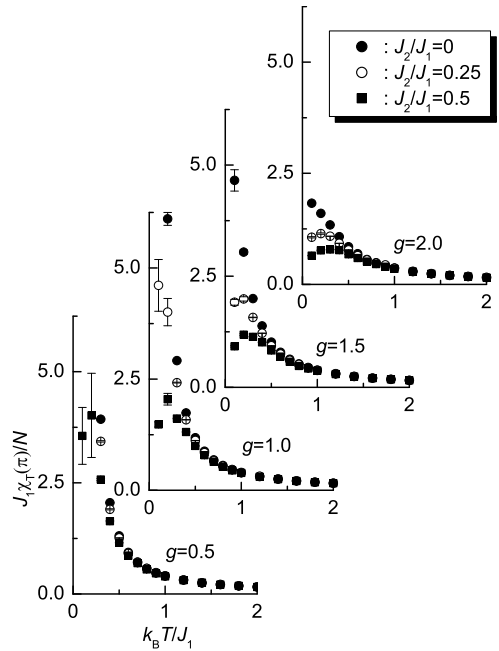}
\caption{Temperature dependence of the staggered susceptibility 
for pseudospins $\chi_{\rm T}(\pi)$ calculated for the coupled 
spin-pseudospin Hamiltonian.}
\end{center}
\label{fig:1}
\end{figure}
The response function is defined as 
\begin{equation}
\chi_{ij}=\int_0^\beta\!{\rm d}\lambda\,
\big(\langle S_j^z(-i\lambda)S_i^z\rangle
-\langle S_j^z\rangle\langle S_i^z\rangle\big)
\end{equation}
where $S_j^z(-i\lambda)$ is the Heisenberg representation 
of $S_j^z$ and $\langle\cdots\rangle$ is the canonical 
average.  $\chi_{ij}$ is Fourier transformed to the 
$q$-dependent susceptibility $\chi(q)$, which we calculate 
by the QMC method; the $q\rightarrow 0$ limit gives the 
uniform spin susceptibility $\chi(T)$ and the staggered 
susceptibility is defined as $\chi(q)$ at $q=\pi$.  
In the following, we use the subscripts ${\rm S}$ 
and ${\rm T}$ as in $\chi_{\rm S}(q)$ and 
$\chi_{\rm T}(q)$, which stand for the spin and 
pseudospin degrees of freedom, respectively.  

The phase diagram of the quantum Ising model ${\cal H}_0$ 
is well known;\cite{sachdev} at $T=0$ there is a long-range 
order for $g<1$ ($g=1$ is a quantum critical point), which 
corresponds to the zigzag (or `antiferromagnetic') CO.  
The calculated staggered susceptibility for pseudospins 
is shown in Fig.~1, where we find that it shows divergent 
behavior at $T\rightarrow 0$ for $g<1$.  
The dispersion relation of the pseudospin excitation 
observed in the calculated dynamical structure factor 
(shown in Fig.~2, see below) agrees well with the 
exact result:\cite{sachdev}  
\begin{equation}
\omega_q=\frac{J_1}{2}\sqrt{1+g^2+2g\cos q}.
\end{equation}
We find in Fig.~1 that the inclusion of the 
coupling term ${\cal H}_{\rm ST}$, which introduces the 
quantum fluctuation via the factor $T_i^+T_j^-$, suppresses 
the divergence.  Thus, we may say that the inclusion of the 
spin degrees of freedom in the quantum Ising model for pseudospins 
leads to the unstable long-range CO.  

\subsection{Spin and pseudospin excitation spectra}

\begin{fullfigure}[t]
\vspace{5pt}
\begin{center}
\includegraphics[width=11.0cm,clip]{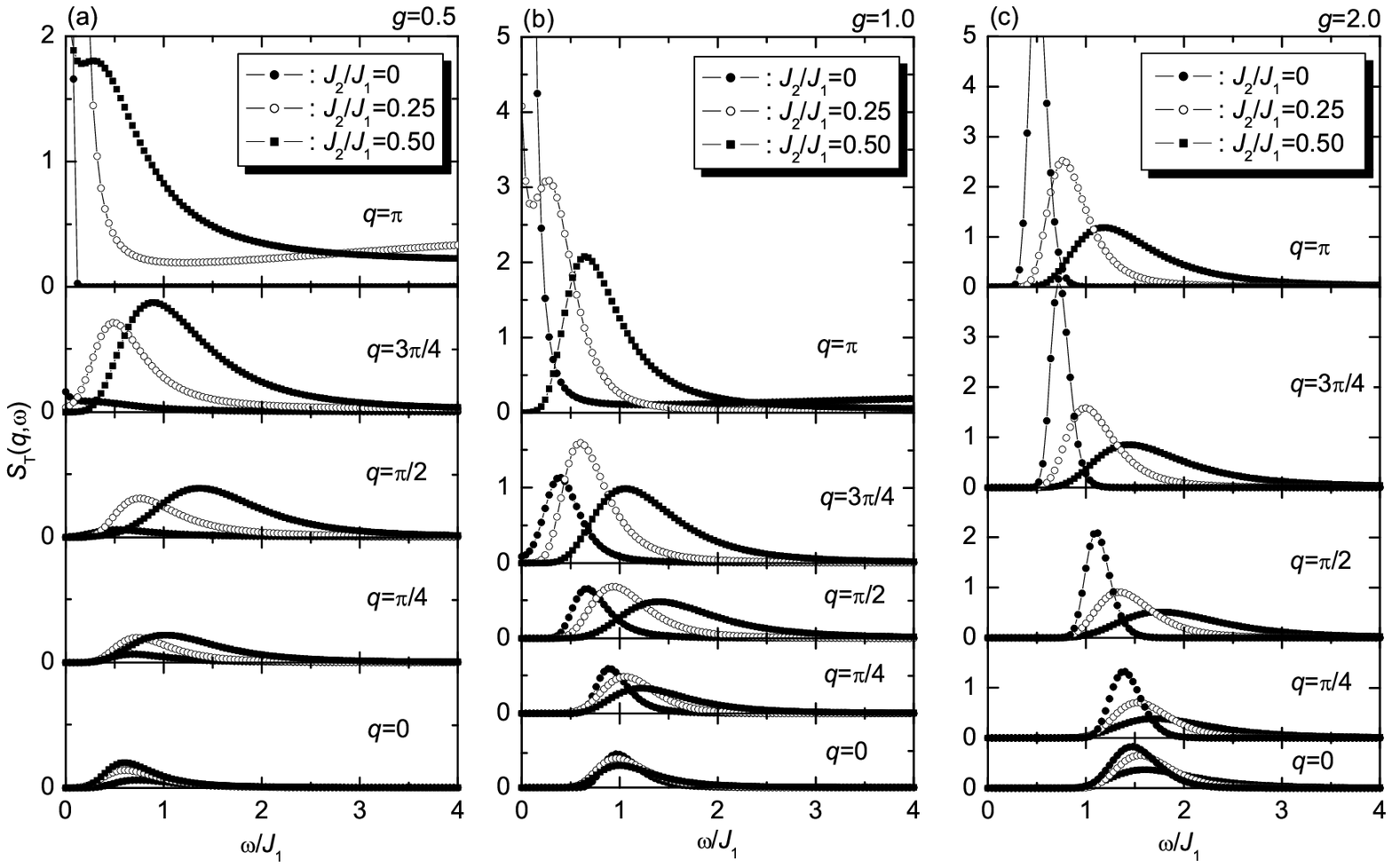}
\caption{Dynamical pseudospin structure factor 
$S_{\rm T}(q,\omega)$ for the coupled spin-pseudospin 
model calculated at $k_{\rm B}T=0.1J_2$.  
The results at $J_2/J_1=0$ are for the quantum Ising 
model.  The peak at $\omega=0$ for $J_2/J_1>0$ 
in the uppermost panel of (a) and (b) is spurious, which 
is due to the error of the maximum entropy method.}
\end{center}
\label{fig:2}
\end{fullfigure}
\begin{fullfigure}[t]
\vspace{5pt}
\begin{center}
\includegraphics[width=11.0cm,clip]{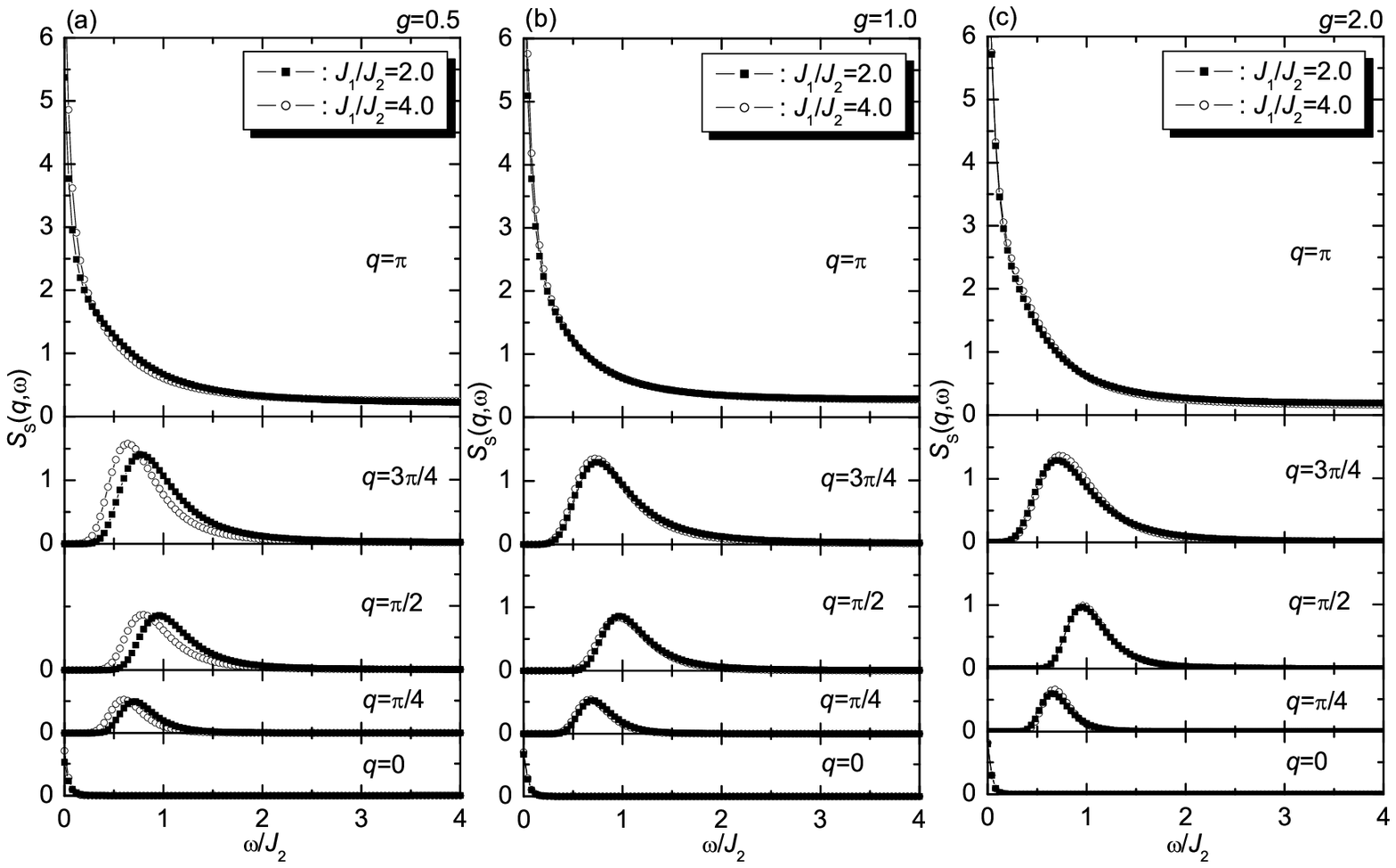}
\caption{Dynamical spin structure factor 
$S_{\rm S}(q,\omega)$ for the coupled 
spin-pseudospin model calculated at 
$k_{\rm B}T=0.1J_2$.}
\end{center}
\label{fig:3}
\end{fullfigure}
The dynamical pseudospin structure factor $S_{\rm T}(q,\omega)$ 
is defined as 
\begin{eqnarray}
&&S_{\rm T}(q,\tau)=\frac{1}{N}\sum_{r_1,r_2}e^{-iq(r_2-r_1)}
\langle T_{r_1}^z(\tau)T_{r_2}^z(0)\rangle\\
&&S_{\rm T}(q,\tau)=\frac{1}{\pi}\int_0^\infty\!{\rm d}
\omega\,S_{\rm T}(q,\omega)K(\omega,\tau)\\
&&K(\omega,\tau)=e^{-\omega\tau}+e^{-\omega(\beta-\tau)}
\end{eqnarray}
where $S_{\rm T}(q,\tau)$ is the Fourier transform of the 
imaginary-time correlation function.  We use the maximum 
entropy method for the inverse Laplace transformation (or 
analytical continuation) to obtain $S_{\rm T}(q,\omega)$ 
from $S_{\rm T}(q,\tau)$.  
The dynamical spin structure factor $S_{\rm S}(q,\omega)$ is 
similarly defined by replacing the pseudospin operator $T^z_r$ 
with the spin operator $S^z_r$.  

The calculated results for the pseudospin excitation spectra 
at low temperature ($k_{\rm B}T=0.1J_2$) are shown in Fig.~2, 
where we find that the spectra are under strong influence of 
the spin-pseudospin coupling term $J_2$.  
With increasing the coupling strength $J_2/J_1$, the peak of 
the pseudospin spectra shifts to higher energies and 
simultaneously the spectra are broadened.  
Thus, the lower-energy edge of the peak is not affected 
strongly by the coupling strength $J_2$, at least when $g$ 
is large.  It seems reasonable to suppose that the scattering 
of the pseudospin excitations due to spin excitations causes 
the broadening of the spectra.  

The calculated results for the spin excitation spectra at 
low temperature are shown in Fig.~3, where we find that, 
in contrast to the pseudospin spectra, the spin excitation 
spectra change very little; i.e., the peak position, width, 
as well as the shape of the spectra are not affected by 
the parameter $J_1$ when $g\gtrsim 1$.  When $g$ is small, 
however, the peak position is slightly shifted to lower 
energies with increasing the value of $J_1$ (see Fig.~3 (a)).  
\begin{figure}[t]
\vspace{5pt}
\begin{center}
\includegraphics[width=7.0cm,clip]{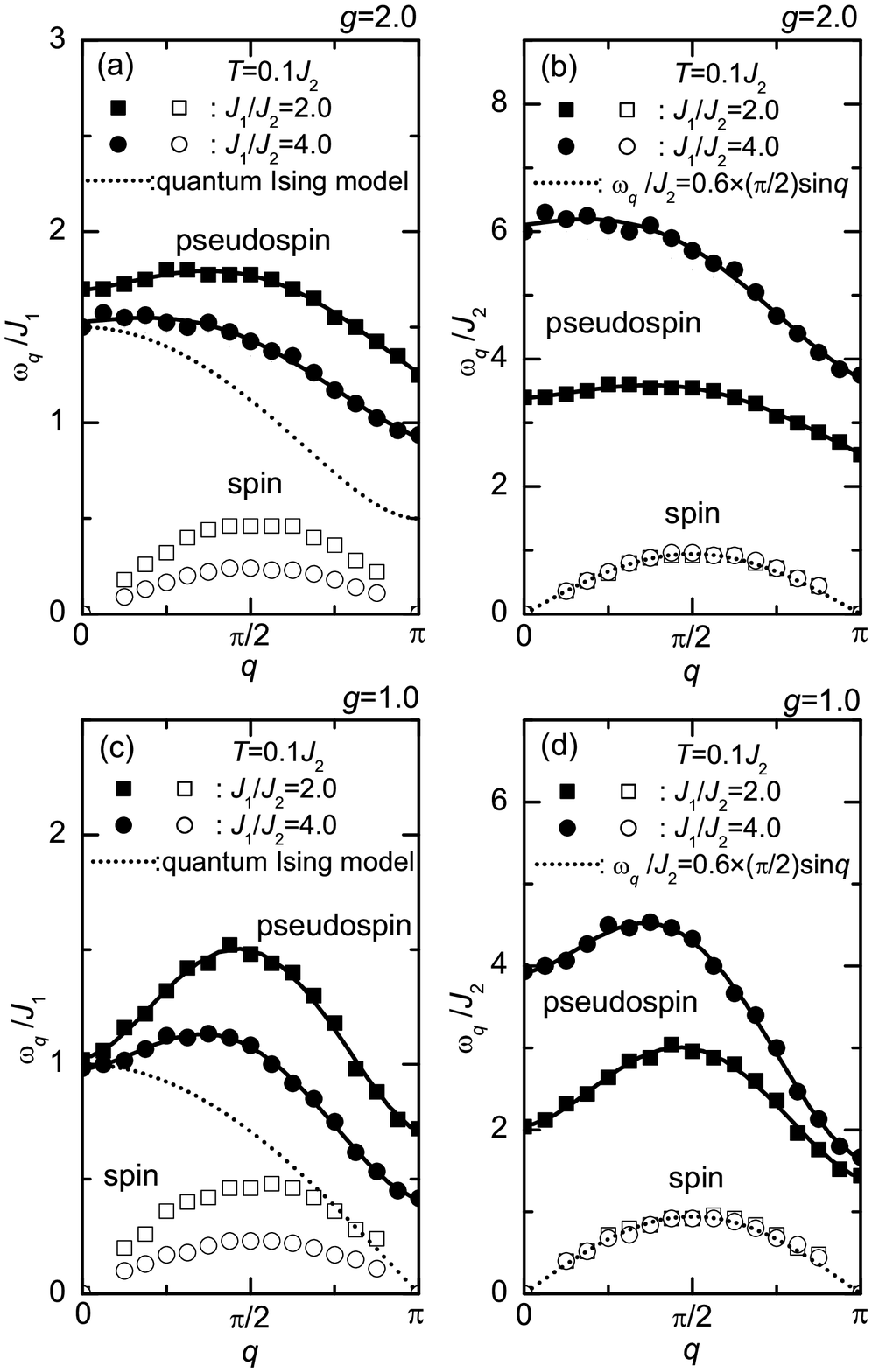}
\caption{Dispersion relations of the spin (open symbols) 
and pseudospin (solid symbols) excitations calculated 
at $k_{\rm B}T=0.1J_2$.  Note that the same data at 
$g=2$ (at $g=1$) are plotted in (a) and (b) (in (c) and 
(d)) in different energy scales $J_1$ and $J_2$.  
The dotted line in (a) and (c) is the dispersion 
relation for the quantum Ising model eq.~(5), and 
that in (b) and (d) is the scaled dispersion 
relation for the 1D antiferromagnetic Heisenberg 
model eq.~(9).}
\end{center}
\label{fig:4}
\end{figure}

The dispersion relation of the spin and pseudospin excitations 
calculated at low temperature are summarized in Fig.~4, which 
are obtained as the momentum dependence of the peak position 
of the spectra.  
For comparison, we show the dispersion of the quantum Ising 
model in Fig.~3 (a) and (c); the gap opens when $g>1$, which 
is closed at $q=\pi$ when $g\rightarrow 1$, leading to the 
`antiferromagnetic' long-range order (or zigzag CO).  
We note that the gap remains open irrespective of the value 
of $g$ when we include the coupling term $J_2$.  
In Fig.~4, we present the same dispersion relations in a different 
energy scales, i.e., $\omega_q/J_1$ and $\omega_q/J_2$.  
We find that, unless $g$ is small, the spin excitation spectra 
are always {\em inside} the charge gap, i.e., inside the gap 
of the pseudospin excitation spectrum; when the charge gap is 
large, the energy scale of the spin excitations is separated 
from the high-energy charge excitations.  With decreasing $g$, 
however, the energy of the charge excitation decreases at the 
momentum $q=\pi$ to couple with the spin excitations.  
We find in Fig.~4 (b) and (d) that for $g\gtrsim 1$ the dispersion 
of the spin excitation spectra scales very well with $J_2$; i.e., 
it does not depend on the value of $J_1$.  
The dispersion of the calculated spin excitation spectra is 
fitted well with the dispersion of the 1D antiferromagnetic 
Heisenberg model 
\begin{equation}
\omega_q/J_2=0.6\times\frac{\pi}{2}\sin q
\end{equation}
if we include the factor $0.6$ as in eq.~(9).  
The factor is independent of $J_1$ for $g\gtrsim 1$ and 
at low $T$.  

These results suggest that at low temperatures there is a 
parameter region where the spin degrees of freedom 
behaves independently from the pseudospin degrees of 
freedom; it is when $g\gtrsim 1$ and the gap of the pseudospin 
excitation spectra is large, inside of which there is 
a spin excitation spectra.  
Thus, we suggest the validity of the decoupling of 
the coupling term of the Hamiltonian as 
\begin{equation}
{\cal H}_{\rm ST}\Rightarrow J_2\sum_i
\big<T_i^+T_{i+1}^-+{\rm H.c.}\big>
\big({\bf S}_i\cdot{\bf S}_{i+1}-\frac{1}{4}\big)
\end{equation}
with
\begin{equation}
\big<T_i^+T_{i+1}^-+{\rm H.c.}\big>\simeq 0.6
\end{equation}
which leads to the effective Heisenberg-model description 
of the spin degrees of freedom of our model.  

\subsection{Uniform spin susceptibility}

\begin{figure}[t]
\vspace{5pt}
\begin{center}
\includegraphics[width=8.5cm,clip]{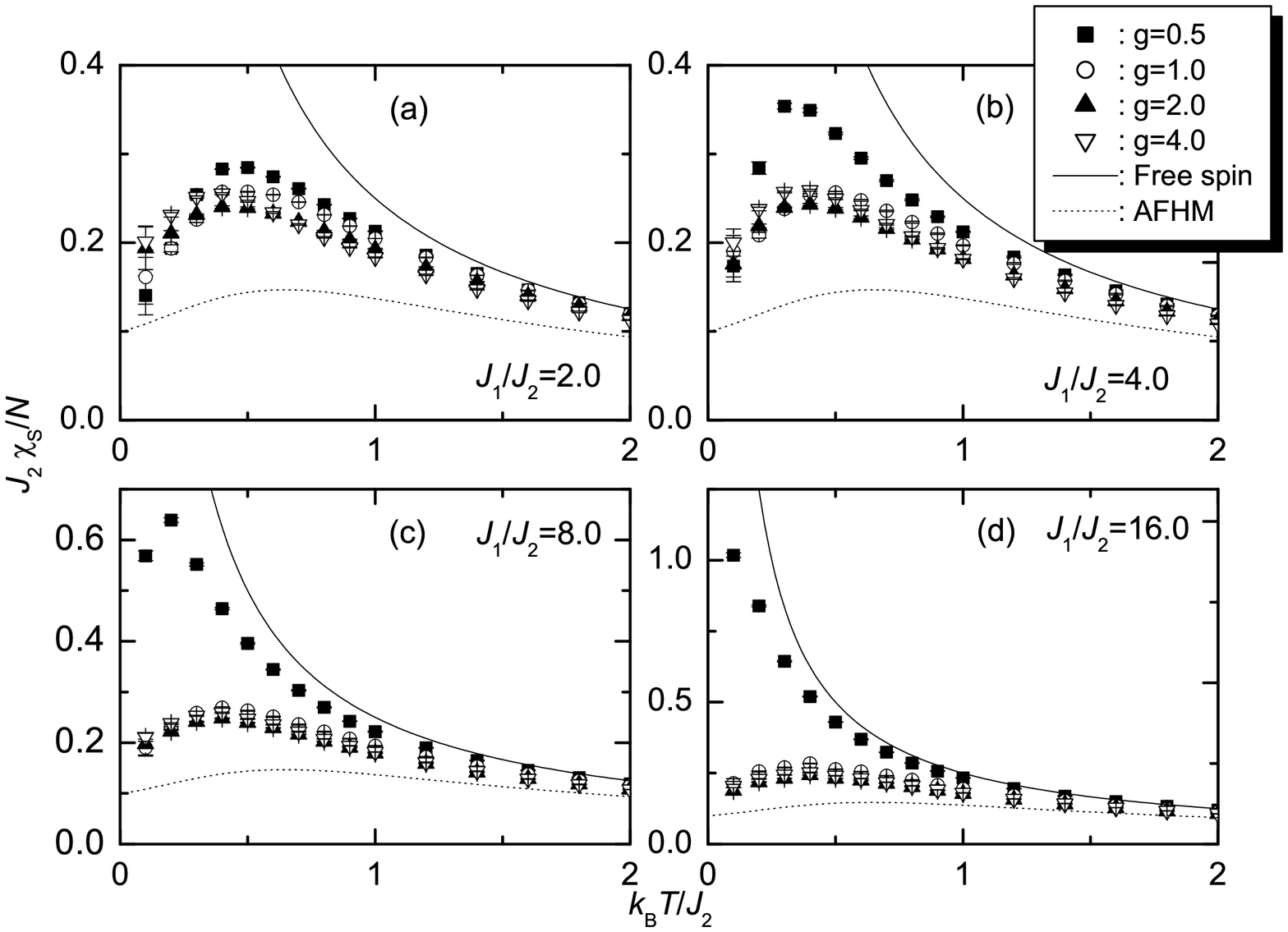}
\caption{Temperature dependence of the uniform spin susceptibility 
$\chi_{\rm S}$ calculated for the coupled spin-pseudospin Hamiltonian.  
The solid and dotted curves are the uniform susceptibility for 
the system of noninteracting $S=1/2$ spins and that for the 1D 
antiferromagnetic Heisenberg model, respectively.}
\end{center}
\label{fig:5}
\end{figure}
To see the validity of the effective Heisenberg-model 
description further, in particular for its temperature 
dependence, we calculate the temperature dependence of 
the uniform spin susceptibility for the coupled 
spin-pseudospin Hamiltonian.  
The results are shown in Fig.~5, where comparisons are 
made with the uniform susceptibility for the system of 
free spins and with that for the 1D antiferromagnetic 
Heisenberg model.  
We find that the temperature $k_{\rm B}T/J_2$ at which 
$J_2\chi_{\rm S}(T)$ shows a maximum is lower than that 
of the 1D antiferromagnetic Heisenberg model; it becomes 
lower with decreasing the value of $g$ or with increasing 
the value of $J_1/J_2$.  In other words, the deviation 
from the Heisenberg model is large when the quantum 
fluctuation of the pseudospins is small, which occurs 
when $g$ is small or $J_1$ is large.  

Now, let us analyze the data more precisely.  In order to 
do this, we fit the results with the temperature dependence 
of the spin susceptibility of the 1D antiferromagnetic 
Heisenberg model, the so-called Bonner-Fisher 
curve;\cite{bonner} i.e., we introduce the $T$-dependent 
{\em effective} exchange coupling constant $J_{\rm eff}(T)$ 
and we determine the values so as to fit the calculated 
uniform spin susceptibility $\chi_{\rm S}(T)$.  
If the values of $J_{\rm eff}$ thus obtained do not depend 
on $T$, it follows that the spin degrees freedom of our 
spin-pseudospin model is reduced to a 1D Heisenberg model 
\begin{equation}
{\cal H}_{\rm spin}=J_{\rm eff}\sum_i\big({\bf S}_i
\cdot{\bf S}_{i+1}-\frac{1}{4}\big)
\end{equation}
at least for the response to the uniform magnetic field.  
The results are shown in Fig.~6.  
We find that the estimated value of $J_{\rm eff}(T)$ 
is indeed a constant for temperatures below 
$k_{\rm B}T\lesssim 0.7J_2$ at $g=2$.  A crossover 
temperature $T^*$ $(=0.7J_2)$ is thereby defined.  
The effective exchange coupling constant takes a value 
\begin{equation}
J_{\rm eff}\simeq 0.6J_2
\end{equation}
which is consistent with the value estimated from the 
dispersion relation of the spin excitation spectra 
(see \S3.2).  We find that also at $g=4$ the scaling behavior 
holds up to a higher temperature ($k_{\rm B}T\lesssim 0.8J_2$), 
but with the same value of $J_{\rm eff}$ (see Fig.~6 (d)), 
demonstrating the validity of the effective Heisenberg-model 
description at $T>T^*$.  At $g=1$, however, the temperature 
region where $J_{\rm eff}(T)$ takes a constant value is 
already very small, although the value is still 
$J_{\rm eff}\sim 0.6J_2$ at $T\sim 0$ K, and at $g=0.5$, 
the value of $J_{\rm eff}$ at $T\sim 0$ K deviates largely 
from $J_{\rm eff}=0.6J_2$ (or decreases strongly when 
$J_1/J_2$ is large), where the effective Heisenberg-model 
description completely fails.  
\begin{figure}[t]
\vspace{5pt}
\begin{center}
\includegraphics[width=8.5cm,clip]{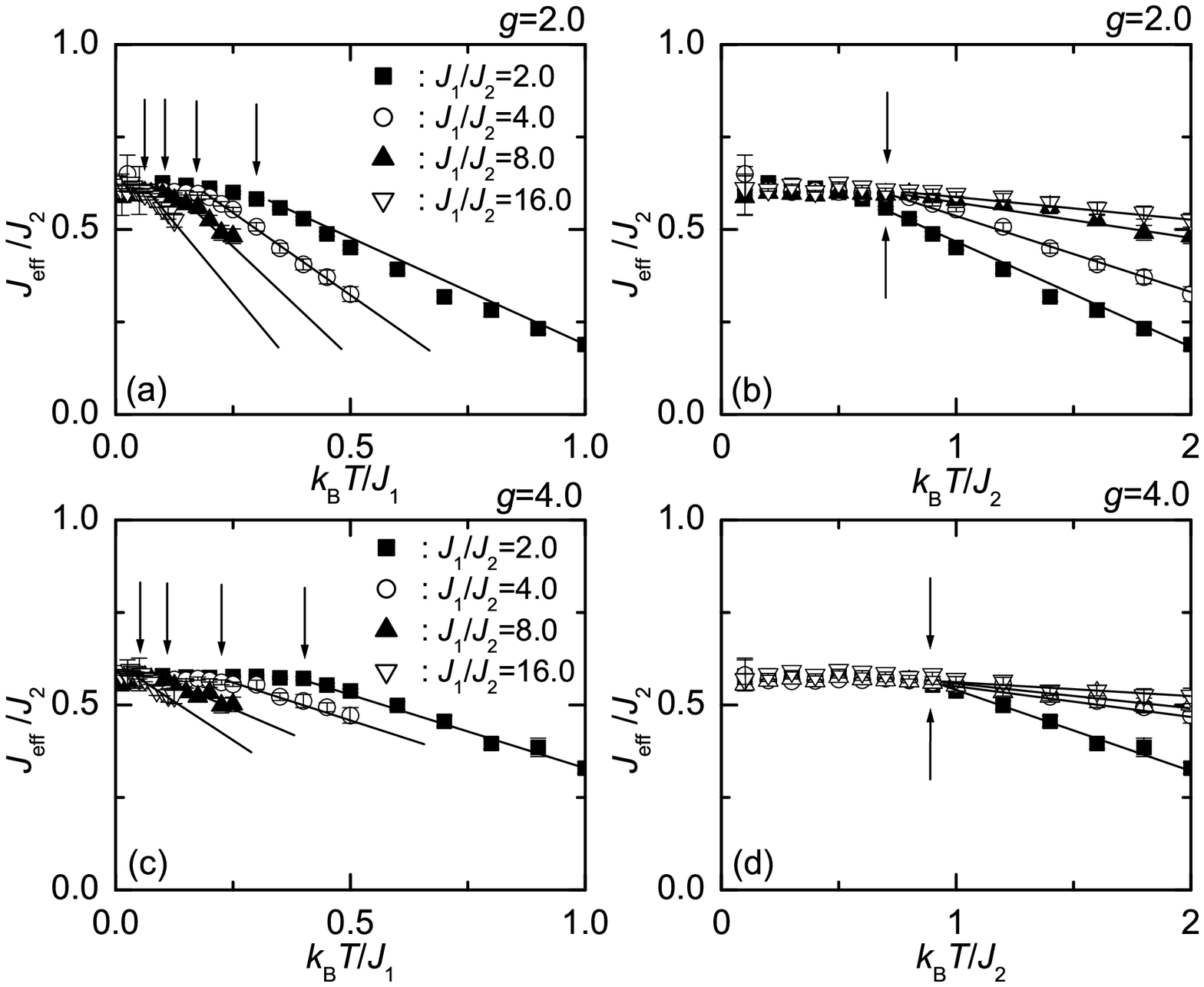}
\caption{Effective exchange coupling constant $J_{\rm eff}(T)$ 
estimated from the fitting of the calculated uniform spin 
susceptibility to the Bonner-Fisher curve.\cite{bonner}  
Note that the same data are plotted as a function of 
$k_{\rm B}T/J_1$ (left panels) and of $k_{\rm B}T/J_2$ 
(right panels), whereby a scaling behavior is seen in the 
latter.  The arrows indicate the crossover temperature $T^*$.  
The solid lines are the guide to the eye.}
\end{center}
\label{fig:6}
\end{figure}

We note here that the crossover temperature $T^*$ roughly 
scales with $J_2$ rather than $J_1$, as seen in Fig.~6.  
One might suppose that it should scale with the size of 
the charge gap: i.e., up to temperatures corresponding 
to the energy of the lowest charge excitations, with which 
the pseudospins can excite, the spin excitations may be 
written in terms of the 1D antiferromagnetic Heisenberg 
model.  However, as we have discussed in \S3.2, the size 
of the charge gap shows a rather complicated behavior and 
does not simply scale with either $J_2$ or $J_1$.  The naive 
picture thus does not hold.  It may be said however that, 
since there is no other excitations available, the 
deviation from the 1D Heisenberg-model description is 
necessarily due to the pseudospin excitations.  

\section{Summary and Discussion}

We have calculated the spin and pseudospin excitation 
spectra and the temperature dependence of the uniform 
spin susceptibility of the coupled spin-pseudospin 
Hamiltonian by using the QMC method.  
We have first shown that, when the pseudospin quantum 
fluctuation is large ($g\gtrsim 1$), the dispersion 
relation of the spin exitation spectra of our model at 
low temperatures agrees well with that of the 1D 
antiferromagnetic Heisenberg model with the renormalized 
effective exchange coupling constant $J_{\rm eff}=0.6J_2$ 
that is independent of the energy scale of the pseudospin 
system $J_1$.  Here, the spin excitation spectra is 
well inside the charge gap, and thus the spin degrees 
of freedom are separated from the charge degrees of 
freedom.  
We then have shown that the temperature dependence of 
the uniform spin susceptibility of our model is well 
described again by the 1D antiferromagnetic Heisenberg 
model with the same effective exchange coupling constant 
$J_{\rm eff}=0.6J_2$.  The description is valid up to 
the crossover temperature $T^*$ that is related to the 
pseudospin excitations of the system and roughly scales 
with $J_2$ unless the quantum fluctuation of the 
pseudospins is small ($g\lesssim 1$).  
We have thus demonstrated the validity of the effective 
Heisenberg-model description of the coupled spin-pseudospin 
model for the quarter-filled ladders.  
It then follows that the coupling between the spin 
and pseudospin degrees of freedom, which occurs at 
$g\lesssim 1$, leads to the {\em anomalous} spin and charge 
dynamics of the system.  

Although the real material $\alpha'$-NaV$_2$O$_5$ 
is modeled well as a 2D trellis-lattice system 
rather than a 1D ladder system and thus we need 
great caution in the direct application of the 
present results, it may be interesting to have a 
rough idea of the values of the physical parameters 
appropriate for $\alpha'$-NaV$_2$O$_5$; 
according to ref.~\cite{nishimoto1}, we have 
$t_\parallel\sim 0.14$ eV, 
$t_\perp\sim 0.30$ eV, 
and $V_\parallel\sim V_\perp\sim 0.8$ eV, 
which lead to 
$J_1\sim 1.6$ eV, 
$J_2\sim 0.10$ eV, and 
$g\sim 0.75$.  
Thus, the real material may be in the region of 
$g\lesssim 1$, where the spin degrees of freedom 
are not separated from the charge degrees of 
freedom.  The anomalous response of the spin 
degrees of freedom may therefore be expected.  
We would here point out, e.g., that the value of 
$J_{\rm eff}$ estimated from the uniform 
susceptibility observed in experiment (which 
takes the value $\sim 600-700$ K at $T\sim 0$ K) 
decreases with increasing temperature,\cite{johnston} 
which is consistent with the results of our 
calculation.  The reported\cite{ohama3} anomalous 
temperature dependence of the nuclear spin-lattice 
relaxation rate $1/T_1$ is also interesting in this 
respect.  
To clarify the dynamics of the spin-charge coupled 
systems near the real CO phase transition, we however 
need not only to examine the region $g\lesssim 1$ 
in greater detail but also to include the 2D coupling 
in the present model, which we want to leave for 
future study.  

Because the anomalous charge dynamics has been noticed 
also in other transition-metal oxides\cite{amasaki} 
and some organic systems\cite{shibata}, we hope that 
the present study will stimulate further researches on 
the intriguing interplay between the spin and charge 
degrees of freedom of strongly correlated electron 
systems with CO instability.  

\section*{Acknowledgements}
We would like to thank A. W. Sandvik, T. Mutou, and 
T. Suzuki for useful discussions on the numerical 
techniques and T. Ohama for enlightening discussion 
on the experimental aspects.  
This work was supported in part by Grants-in-Aid for 
Scientific Research (Nos.~11640335 and 12046216) from 
the Ministry of Education, Culture, Sports, Science, 
and Technology of Japan.  
Computations were carried out at the computer centers 
of the Institute for Molecular Science, Okazaki, and 
the Institute for Solid State Physics, University of 
Tokyo.


\begin{thebibliography}{99}

\bibitem{smolinski} H. Smolinski, C. Gros, W. Weber, 
U. Pechert, G. Roth, M. Weiden, C. Geibel: 
Phys. Rev. Lett. {\bf 80} (1998) 5146.  
\bibitem{seo} H. Seo and H. Fukuyama: 
J. Phys. Soc. Jpn. {\bf 67} (1998) 2602.  
\bibitem{nishimoto1} S. Nishimoto and Y. Ohta: 
J. Phys. Soc. Jpn. {\bf 67} (1998) 2996.  
\bibitem{thalmeier} P. Thalmeier and P. Fulde: 
Europhys. Lett. {\bf 44} (1998) 242.  
\bibitem{mostovoy1} M. V. Mostovoy and D. I. Khomskii: 
Solid State Commun. {\bf 113} (2000) 159. 
\bibitem{isobe} M. Isobe and Y. Ueda: 
J. Phys. Soc. Jpn. {\bf 65} (1996) 1178.  
\bibitem{ohama1} T. Ohama, H. Yasuoka, M. Isobe, 
and Y. Ueda: Phys. Rev. B {59} (1999) 3299.  
\bibitem{sawa} H. Sawa, E. Ninomiya, T. Ohama, H. Nakao, 
K. Ohwada, Y. Murakami, Y. Fujii, Y. Noda, M. Isobe, 
and Y. Ueda: J. Phys. Soc. Jpn. {\bf 71} (2002) 385.  
\bibitem{johnston} D. C. Johnston, R. K. Kremer, M. Troyer, 
X. Wang, A. Kl\"umper, S. L. Bud'ko, A. F. Panchula, 
and P. C. Canfield: Phys. Rev. B {\bf 61} (2000) 9558.  
\bibitem{ohama2} For a review, see T. Ohama: 
Bussei Kenkyu (Kyoto) {\bf 74} (2000) 391 
[in Japanese].  
\bibitem{ravy} S. Ravy, J. Jegoudez, and A. Revcolevschi: 
Phys. Rev. B {\bf 59} (1999) R681.  
\bibitem{nakao} H. Nakao, K. Ohwada, N. Takesue, Y. Fujii, 
M. Isobe, and Y. Ueda: Physica B {\bf 21-243} (1998) 534.  
\bibitem{damascelli} A. Damascelli, C. Presura, 
D. van der Marel, J. Jegoudez, and A. Revcolevschi: 
Phys. Rev. B {\bf 61} (2000) 2535.  
\bibitem{presura} C. Presura, D. van der Marel, 
A. Damascelli, and R. K. Kremer: 
Phys. Rev. B {\bf 61} (2000) 15762.  
\bibitem{nishimoto2} S. Nishimoto and Y. Ohta: 
J. Phys. Soc. Jpn. {\bf 67} (1998) 3679.  
\bibitem{nishimoto3} S. Nishimoto and Y. Ohta: 
J. Phys. Soc. Jpn. {\bf 67} (1998) 4010.  
\bibitem{mostovoy2} M. V. Mostovoy, J. Knoester, 
and D. I. Khomskii: Phys. Rev. B {\bf 65} (2002) 064412.  
\bibitem{hemberger} J. Hemberger, M. Lohmann, M. Nicklas, 
A. Loidl, M. Klemm, G. Obermeier, and S. Horn: 
Europhys. Lett. {\bf 42} (1998) 661.  
\bibitem{ohama3} T. Ohama {\it et al.}: unpublished.  
\bibitem{cuoco} M. Cuoco, P. Horsch, and F. Mack: 
Phys. Rev. B {\bf 60} (1999) R8438.  
\bibitem{sa} D. Sa and C. Gros: 
Eur. Phys. J. B {\bf 18} (2000) 421.  
\bibitem{nakaegawa} For details, see T. Nakaegawa: 
{\it Master thesis} (Chiba University, 2002).  
\bibitem{sachdev} S. Sachdev: {\it Quantum Phase Transitions} 
(University Press, Cambridge, 1999).  
\bibitem{bonner} J. C. Bonner and M. E. Fisher: 
Phys. Rev. {\bf 135} (1964) A640.  
\bibitem{amasaki} R. Amasaki, Y. Shibata, and Y. Ohta: 
cond-mat/0110333.  
\bibitem{shibata} Y. Shibata, S. Nishimoto, and Y. Ohta: 
Phys. Rev. B {\bf 64} (2001) 235107.  

\end{thebibliography}
\end{document}